\documentclass[aps,prl,twocolumn,preprintnumbers,superscriptaddress,10pt,longbibliography]{revtex4-2}
\usepackage{amsmath}  
\usepackage{amssymb}  
\usepackage[utf8]{inputenc}
\usepackage{hyperref}
\usepackage{graphicx}  
\usepackage{color}
\usepackage{ulem}
\newcommand{\eq}{\begin{equation}}
\newcommand{\eqx}{\end{equation}}
\newcommand{\eqn}{\begin{eqnarray}}
\newcommand{\eqnx}{\end{eqnarray}}

\usepackage[utf8]{inputenc}
\usepackage[T1]{fontenc}

\begin{document}
\title{Behavior of the scaling correlation functions under severe subsampling}

\author{Sabrina Camargo}
\email{scamargo@unsam.edu.ar}
\affiliation{Instituto de Ciencias F\'isicas (ICIFI-CONICET), Center for Complex Systems and Brain Sciences (CEMSC3), Escuela de Ciencia y Tecnología, Universidad Nacional de Gral. San Martín, Campus Miguelete, 25 de Mayo y Francia,  1650, San Martín, Buenos Aires, Argentina}
\affiliation{Consejo Nacional de Investigaciones Cient\'{\i}fcas y T\'ecnicas (CONICET), Godoy Cruz 2290, 1425, Buenos Aires, Argentina.}

\author{Nahuel Zamponi}
\email{zamponi.n@gmail.com}
\affiliation{Division of Hematology and Medical Oncology, Department of Medicine, Weill Cornell Medicine, 1300 York Avenue, New York, NY 10065, USA.}

 \author{Daniel A. Martin}
\affiliation{Instituto de Ciencias F\'isicas (ICIFI-CONICET), Center for Complex Systems and Brain Sciences (CEMSC3), Escuela de Ciencia y Tecnología, Universidad Nacional de Gral. San Martín, Campus Miguelete, 25 de Mayo y Francia,  1650, San Martín, Buenos Aires, Argentina}
\affiliation{Consejo Nacional de Investigaciones Cient\'{\i}fcas y T\'ecnicas (CONICET), Godoy Cruz 2290, 1425, Buenos Aires, Argentina.}

\author{Tatyana Turova}   
\affiliation{Mathematical Statistics, University of Lund, Box 118, 221 00 Lund, Sweden}

\author{Tom\'as S.\ Grigera}
\affiliation{Instituto de F\'isica de L\'iquidos y Sistemas Biol\'ogicos (IFLySiB), Universidad Nacional de La Plata, Calle 59 n 789, 1900 La Plata, Argentina}
\affiliation{Departamento de F\'isica, Facultad de Ciencias Exactas,  Universidad Nacional de La Plata, 1900 La Plata, Argentina}
\affiliation{Consejo Nacional de Investigaciones Cient\'{\i}fcas y T\'ecnicas (CONICET), Godoy Cruz 2290, 1425, Buenos Aires, Argentina.}

\affiliation{Istituto dei Sistemi Complessi, Consiglio Nazionale delle Ricerche, via dei Taurini 19, 00185 Rome, Italy} 

\author{Qian-Yuan Tang}
\affiliation{Department of Physics, Faculty of Science, Hong Kong Baptist University, Hong Kong SAR, China }

\author{Dante R. Chialvo}
 \affiliation{Instituto de Ciencias F\'isicas (ICIFI-CONICET), Center for Complex Systems and Brain Sciences (CEMSC3), Escuela de Ciencia y Tecnología, Universidad Nacional de Gral. San Martín, Campus Miguelete, 25 de Mayo y Francia,  1650, San Martín, Buenos Aires, Argentina}
 \affiliation{Department of Physics, Faculty of Science, Hong Kong Baptist University, Hong Kong SAR, China }
\affiliation{Consejo Nacional de Investigaciones Cient\'{\i}fcas y T\'ecnicas (CONICET), Godoy Cruz 2290, 1425, Buenos Aires, Argentina.}


\begin{abstract}
Scale-invariance is a ubiquitous observation in the dynamics of large distributed complex systems. The computation of its scaling exponents, which provide clues on its origin, is often hampered by the limited available sampling data, making an appropriate mathematical description a challenge. 
This work investigates the behavior of correlation functions in fractal systems under conditions of severe subsampling. Analytical and numerical results reveal a striking robustness: the correlation functions continue to capture the expected scaling exponents despite substantial data reduction. This behavior is demonstrated numerically for the random 2-D Cantor set and the Sierpinski gasket, both consistent with exact analytical predictions. Similar robustness is observed in 1-D time series both synthetic and experimental, as well as in high-resolution images of a neuronal structure. Overall, these findings are broadly relevant for the structural characterization of biological systems under realistic sampling constraints.

\end{abstract}

\flushbottom
\maketitle

 The presence of statistical scale-invariant correlations, both in time and space is almost  a hallmark of a complex system \cite{ref_general,Vicsek}  justifying its careful exploration as a first step of the system' analysis. While in numerical models its characterization in the limit of infinite data size and sampling is straightforward, in most experimental applications, including biology, the data is scarce,  the system may be large  and the sampling is sub-optimal. The appropriate  estimation of correlation functions is relevant to understand the system behavior, as for instance in the experimental study of subcellular structures under healthy and pathological disturbances  \cite{zamponi2018,zamponi2022}. The motivation of the present work  is  to clarify how suboptimal sampling of the data may affect the properties of the correlation functions.
  
To that end we analyze the commonly used metrics of scale-invariance, commenting on its performance in estimating characteristic exponents as a function of the sampled fraction.   The remaining of  the paper is organized as follows: In the next paragraphs  the main analytical considerations are discussed. After that  numerical results are described for two well known prototypical sets, namely the random Cantor set and the Sierpinsky gasket. The case of synthetic 1-D time series  is discussed next followed by the analysis of experimental data  to gauge the effects of subsampling. The paper closes with a short discussion of caveats and related results.

The two most common practical approaches for evaluating scale invariant sets are the radial distribution function and the fractal dimension, the latter often computed using the box-counting algorithm.

\textit{Fractal dimension:} 
The fractal dimension \( D_f \) of a set can be estimated  by the box-counting algorithm which analyzes how the number of covering boxes scales with box size \cite{Vicsek}. The steps involve to
 cover the set with a grid of boxes of side length \( \varepsilon \),
and count the number of boxes \( N(\varepsilon) \) that contain part of the set.
The process is repeated for a range of decreasing \( \varepsilon \) values and finally
   the slope of the linear regression line through the  \(\ln N(\varepsilon)\) versus \(\ln (1/\varepsilon)\) points  is computed. The fractal dimension is given by:
\begin{equation}
D_f = \lim_{\varepsilon \to 0} \frac{\ln N(\varepsilon)}{\ln (1/\varepsilon)}. 
\end{equation}
In practice, \( D_f \) is approximated as the negative slope of the best-fit line in the log-log plot:
\begin{equation}
 D_f \approx -\frac{\Delta \ln N(\varepsilon)}{\Delta \ln \varepsilon},
\end{equation}
 for a reasonably small $\varepsilon$.
\begin{figure} [b]
    \centering
    \includegraphics[width=0.98\linewidth]{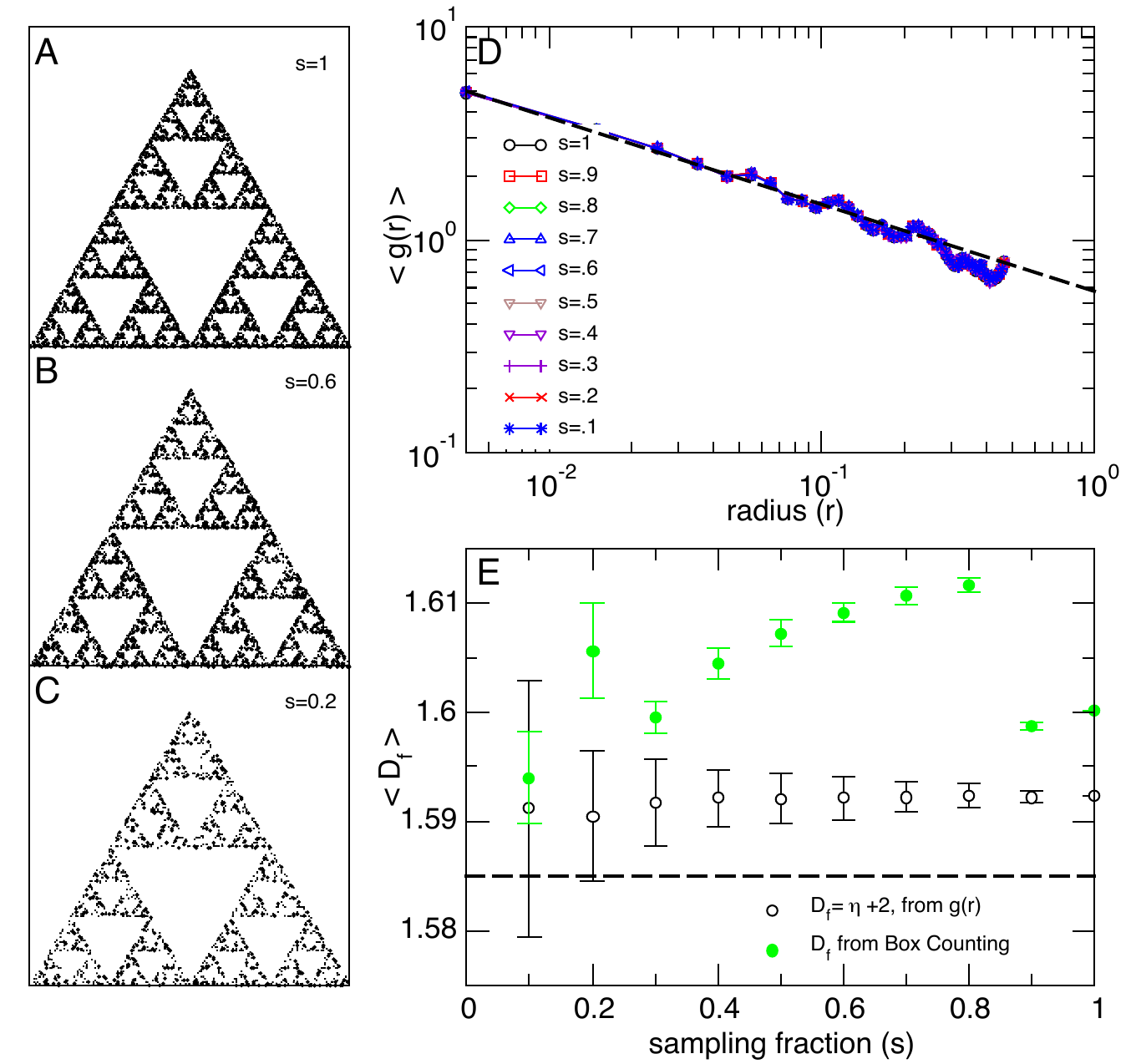}
    \caption{Subsampling of the Sierpinski gasket does not severely affect the numerical estimation of the scaling exponents. A set of 10000 points belonging to the Sierpinski triangle was created by iterating the chaos game rule. Typical sets for full, 0.6 and 0.2 sampling fractions are shown in panels A-C respectively. The results in Panel D show the radial distribution functions $g(r)$ computed for the sampling fractions depicted in the legend, from which the initial power-law decay of $g(r)\sim r^{\eta}$  was computed by a log-linear fit of the estimated $g(r)$.  
    The dashed line correspond to a power law with the exact exponent $\eta = D_f-2=-0.415$ and $D_f = \log(3)/\log(2)$. Panel E illustrates the scaling exponents   vs.\ sampling ratio derived both from the $g(r)$ function as well as from the box counting method. Error bars correspond to mean +/- sd values computed from 100 realizations. The dashed line corresponds to the exact  $D_f =\eta +2=1.585$ value.  }
    \label{fig:Sierp}
\end{figure}

\textit{Radial distribution function:}
The radial distribution function \( g(r) \) is widely used to quantify spatial correlations in systems ranging from  colloids \cite{weeks2000} and granular media \cite{granular} to neuronal networks \cite{camargo,tomasreview} and spatial tree patterns \cite{trees}.
  
For homogeneous systems it can be written as

\begin{equation}
g^{(2)}(\mathbf{r})= \frac{1}{\rho N} \sum_{i\neq j} \left\langle
    \delta[\mathbf{r}-(\mathbf{x}_i-\mathbf{x}_j)]\right\rangle,
\end{equation}

where $\rho=N/V$ is the number density and $\mathbf{x_i}$ are the particles' positions.  The radial distribution function is essentially the two-point density correlation function, normalized to unity at long distances,
\begin{equation}
    g(r) = \frac{1}{\rho^2} \langle \rho(0)\rho(r)\rangle, \qquad r>0,
\end{equation}
where ${\rho(\mathbf{r})}=\sum_i\delta(\mathbf{r}-\mathbf{x}_i)$, and the equality is valid except at $r=0$ where $g(r=0)=0$ and the two-point density correlation is singular.

\textit{Subsampling:}
For each of the structures analyzed below, quantities are computed for the full sample, and then at several subsamplings defined by the probability $s$ of including a given particle of the original structure in the subsample. The quantities are computed exactly in the same way for the original structure and all subsamplings (when involved, the sample size is clearly recomputed for the subsample, i.e.\ $N_s\approx s N$).

\textit{Analytical considerations:}
Given an \(N\)-point snapshot of a system's configuration, the radial distribution function is estimated as (we consider periodic boundary conditions for simplicity)
\begin{equation}
    \hat g(r_k) = \frac{1}{\rho N} \sum_{ij} \frac{\Delta\left[r_{ij} -
               (k+1/2)\delta r \right]}{V_k},    
           \label{eq:gr-estimator}
\end{equation}
where  $\delta r$   is the bin width, $r_k=k \delta r$ is the position of the center of the $k$-th bin and $V_k$ its volume, and $\Delta[r]$ is the interval indicator function
\begin{equation}
  \Delta[r] =
  \begin{cases}
    1  & \text{if } - \delta r/2 < r \le  \delta r/2,\\
    0 & \text{otherwise}.
  \end{cases}
\end{equation}
Defining the number of points in the $k$-th bin centered on particle $i$, $N_i(k)= \sum_j \Delta[r_{ij} - (k+1/2) \delta r]$, we can write the estimator as
\begin{equation}
  \hat g(r_k) = \frac{1}{\rho N} \frac{1}{V_k}  \sum_i N_i(k).
\end{equation}
When the snapshot is subsampled, each point is included with probability $p$  (here $p=s$). Let's call $N_i^{(p)}(k)$ the number of points in the $k$-th bin for a subsampled configuration, given that particle $i$ belongs to the subsample. The $N_i^{(p)}(k)$ are correlated random variables, but the marginal probability of a single of these quantities should be binomial if the subsampling is homogeneous and independent (i.e.\ if the probability of picking a particle is independent of its position and of whether its neighbors have been picked). Then
\begin{equation}
  \left\langle N_i^{(p)}(k) \right\rangle_p = p N_i(k),
\end{equation}
where the average is over all possible subsamplings with probability $p$. The estimate of the radial distribution function from a single subsample is
\begin{equation}
  \hat g^{(p)}(r_k)  = \frac{1}{V_k} \frac{1}{\rho^{(p)} N^{(p)}}   \sideset{}{'}\sum_i{} N_i^{(p)}(k),
  \label{eq:single-subsamp}
\end{equation}
where the sum is primed to remind that the number of terms fluctuates with the subsampling.
 
Averaging over all possible subsamples,
\begin{equation}
 \left\langle  \hat g^{(p)}(r_k) \right\rangle \approx \frac{V}{V_k}  \left\langle \frac{N_i^{(p)}(k)}{N^{(p)} } \right\rangle_p = \hat g(r_k).
\end{equation}
where we have assumed homogeneity (as does Eq.~\ref{eq:gr-estimator}), and the $\approx$ sign is because it is not the single snapshot that is homogeneous, but the ensemble where it comes from.

The main point is that the single-subsample estimate Eq.~\ref{eq:single-subsamp} will actually be a good approximation to the full-sample estimate Eq.~\ref{eq:gr-estimator} as long as $N^{(p)}$ is not too small, because due to double homogeneity of the original sample and of the subsampling, the average over the subsamples can be approximately realized by a space average (average over focal particles), i.e. $\hat g(r_k)$ can be regarded as self-averaging over the ensemble of subsamples.

We can estimate the extent of the fluctuations of $\hat g^{(p)}(r_k)$,

\begin{multline}
  \text{Var}\left[ (\rho^{(p)})^2 \hat g^{(p)}(r_k) \right] 
  = \frac{1}{V^2V_k^2} \Biggl[ \sideset{}{'}\sum_i \text{Var}(N_i^{(p)}) \\
  + \sideset{}{'}\sum_{i\neq j} \left\langle \left( N_i^{(p)} - \left\langle N_i^{(p)}\right\rangle \right) \left( N_j^{(p)} -\left\langle N_j^{(p)}\right\rangle \right) \right\rangle \Biggr].
\end{multline}

Of the $N(N-1)$ terms contributing to the double sum, only a number of order $N$ will be different from zero since  {$N_i^{(p)}$ and $N_j^{(p)}$} become decorrelated when the centers are far apart. Since $\text{Var}\left[N_i^{(p)}(k)\right] = p(1-p) N_i(k)$, we have
\begin{equation}
  \text{Var}\left[ \hat g^{(p)}(r_k) \right] \approx
  \frac{1-p}{N p} \frac{\hat g(r_k)}{V_k \rho} + O\left(\frac{1}{N}\right) ,
\end{equation}
which confirms that $\hat g(r_k)$ is self-averaging for $V\to\infty$.
 
Notably, the estimator $\hat{g}(r_k)$ is computed by binning all pairwise distances into intervals of width  $\delta r$, so it can be formally written as a convolution of the exact radial distribution function $g(r)$ with a rectangular kernel:
\begin{equation}
    \hat{g}(r_k) = \int g(r) \, K_{\delta r}(r_k - r) \, \mathrm{d}r,
\end{equation}
where $K_{ \delta r}(r) = \frac{1}{ \delta r} \, \Delta(r)$ is the normalized box function centered at zero. In Fourier space, this becomes:
\begin{equation}
    \widehat{\hat{g}}(q) = \widehat{g}(q) \cdot \mathrm{sinc}(q   \delta r / 2),
\end{equation}
where $q$ denotes the spatial frequency (wavenumber) dual to $r$. This expression makes the low-pass filtering effect explicit: the sinc-shaped transfer function attenuates high-frequency components of $g(r)$, effectively imposing a spatial frequency cutoff near $1/ \delta r$. As a result, the estimator suppresses small-scale fluctuations while preserving the low-frequency content that governs the scaling behavior.

Now we turn to describe numerical results of the effect of subsampling synthetic fractal structures using two well-known fractal sets where the fractal dimension is known analytically.
\begin{figure} [ht!]
    \centering
    \includegraphics[width=0.98\linewidth]{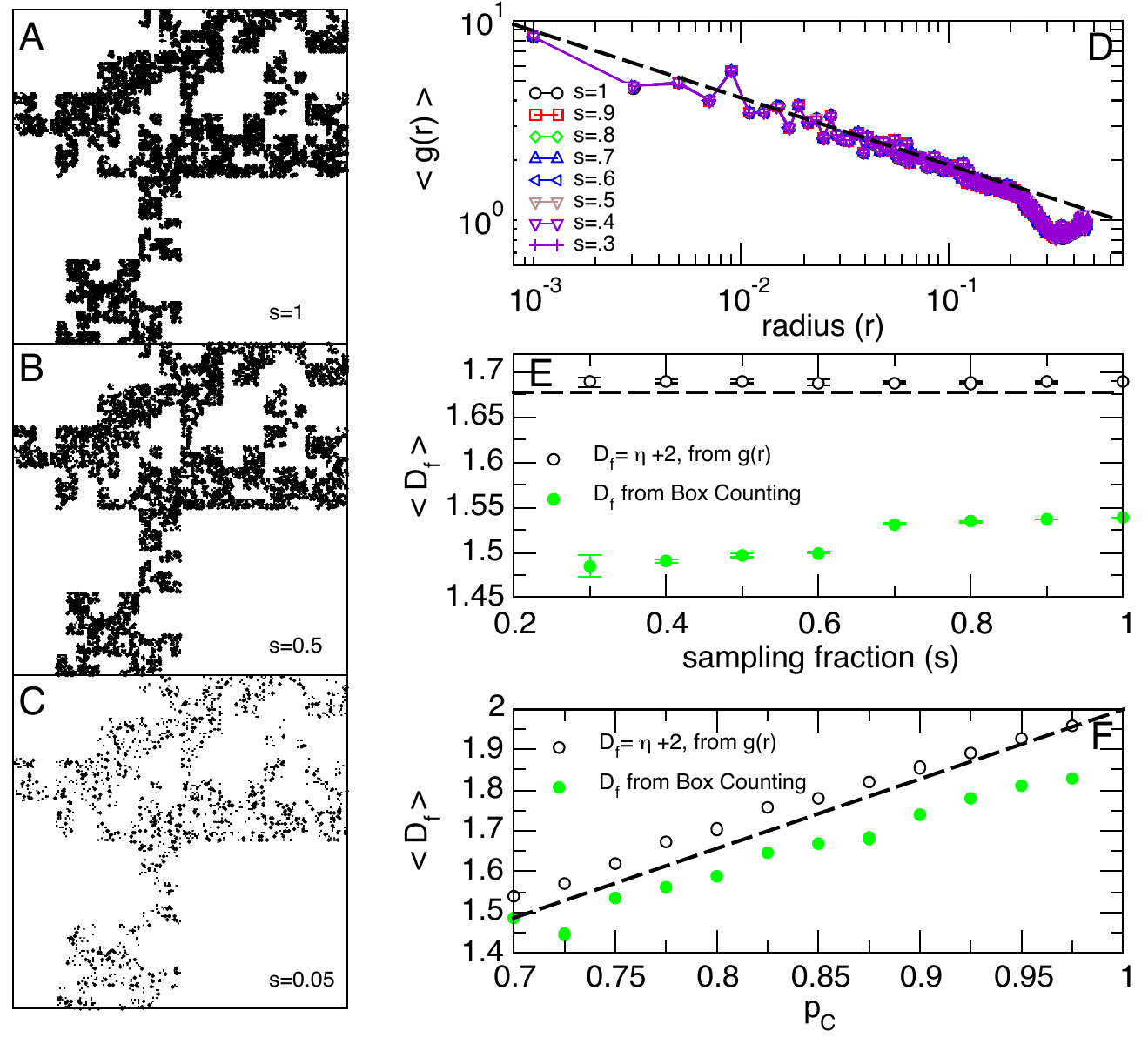}
    \caption{Numerical estimation of the scaling exponents of the 2-D random Cantor set under a wide range of subsampling. Examples of the Cantor sets (for $p_C=0.8$) are shown in Panel A (fully sampled) as well as two subsampled versions (s=0.5 and 0.05) in Panels B and C respectively.  
Panel D: radial distribution functions $g(r)$ for $p_C=0.8$ at different subsamplings.  
 Panel E: Fractal dimension estimated from the $g(r)$ and from the box counting computations of the 2-D random Cantor set with $p_C = 0.8$ at different sampling fractions. Panel F: Same estimations as in Panel E for the fully sampled 2-D random Cantor set generated with different values of $p_C$. Statistics  (means +/- standard deviation in Panels D and E, means in Panel F) computed from 10 independent realizations, dashed lines in panels D-F correspond to the analytical scaling values. }
 \label{fig:Cantor}
\end{figure}
 
\textit{The Sierpinski gasket:}
The Sierpinski gasket, or Sierpinski triangle, is a self-similar set with fractal dimension $D_f=\log_2 3$. It can be built in several ways, here we have used the so-called chaos game\cite{IFS,IFS2} where starting from a point that belongs to the set, one obtains another point by moving half-way towards a randomly selected vertex of the equilateral triangle containing the set. Specifically, taking an equilateral triangle defined by the vertices $\mathbf{v}_1 = (0,0)$, $\mathbf{v}_2 = (1,0)$, $\mathbf{v}_3 = (1/2,\sqrt{3}/2)$, and setting $\mathbf{p}_1=\mathbf{v}_1$, successive points belonging to the fractal are obtained by the random sequence 
\begin{equation}
    \mathbf{p}_{n+1} = \frac{1}{2} \left( \mathbf{p}_n + \mathbf{v}_r\right),
\end{equation}  
where $r$ is a random integer between 1 and 3.

The results for the Sierpinski gasket are shown in Fig~\ref{fig:Sierp} depicting the $g(r)$ and the initial decay exponent for the full sample of the fractal and for several subsamplings. Notice that the different curves in Fig~\ref{fig:Sierp} are almost indistinguishable from each other. The theoretical value of $\eta$ is $\eta = D_f-2 = \log3/\log2-2 \approx -0.415$, a numerical estimate of $D_f$ which is within 0.5\% of the theoretical value, and very stable against (even severe) subsampling. In passing, we note that the correlation estimation based on box counting are less stable than $g(r)$, an observation also made with other sets commented later on.

  \begin{figure} [ht!]
    \centering
    \includegraphics[width=1\linewidth]{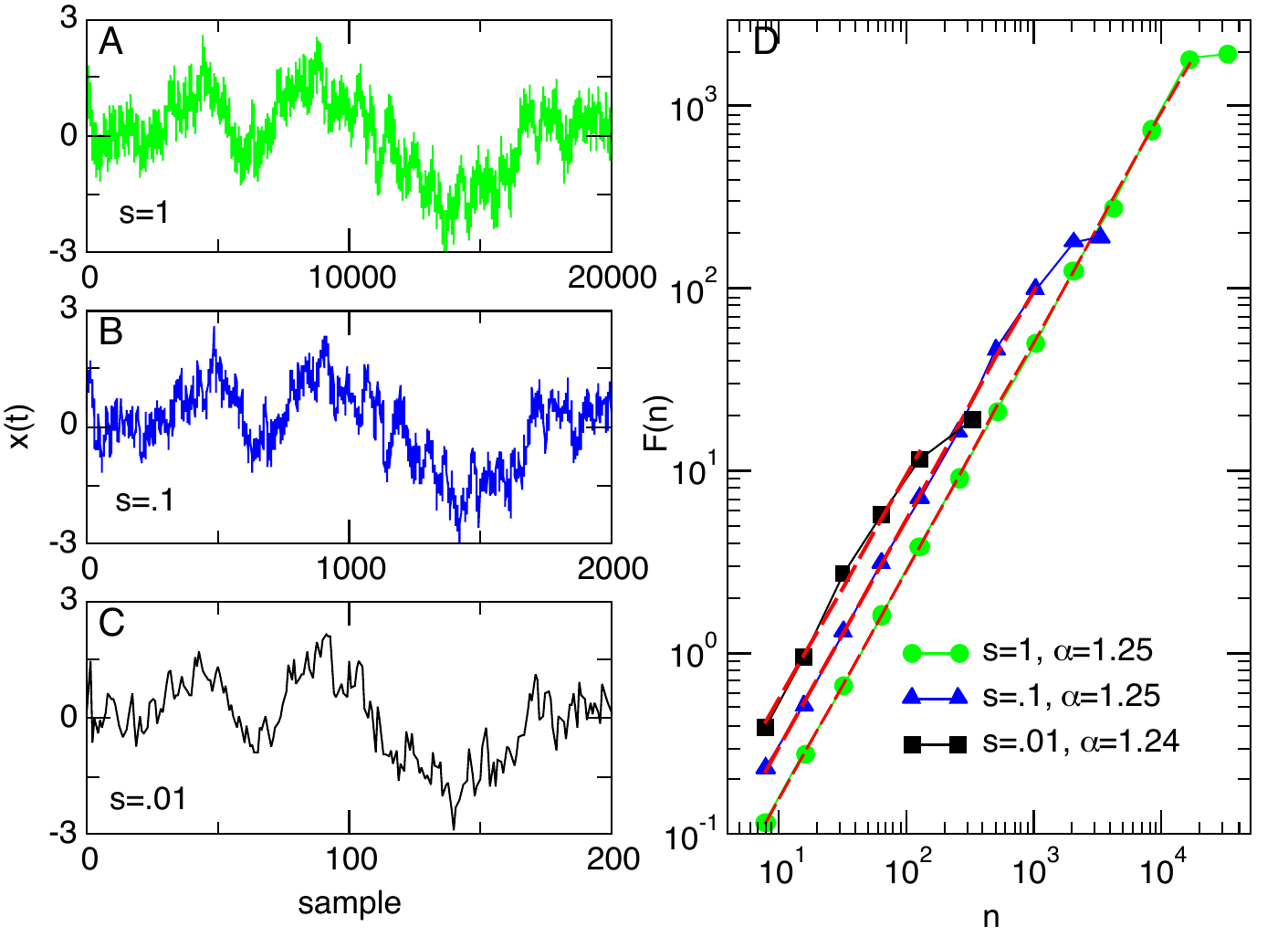}
    \caption{Time series long range correlations behavior under random  subsampling. Panel  A illustrates a typical example of a Gaussian correlated time series with $1/f^{\beta}$ spectral decay ($\beta=1.5$, $\sigma^2 = 1$). Panels B and C depict the  subsampled versions at  $s=0.1$, and s=$0.01$ respectively. Panel D shows the detrended fluctuation analysis of each time series, where $F(n)$ is the total fluctuation computed from  segments of length  $n$, and a behavior $F\propto n^\alpha$ is expected. The computed value of the $\alpha$ scaling exponent remains very close to the predicted dependency on the spectral $\beta$, i.e., $\alpha=(\beta+1)/2 =1.25$, even for the case of extreme subsampling. Note that the subsampling seems to remove high frequency fluctuations, thus limiting the  scaling regions to smaller $n$ values (calculation done with a fully sampled ($s=1$) time series of $N=2^{15}$).}
    \label{fig:1D}
\end{figure}
 
\textit{The 2-D random  Cantor set:}
 We also consider the random Cantor set \cite{IFS}. Its stochastic character entails the concept of the almost sure Hausdorff dimension, meaning it holds with probability 1 for a random realization of the set.
The random Cantor set in 2-D is constructed through an iterative process that generalizes the 1-D random Cantor set to two dimensions. Starting with a unit square (or any initial square) in 2-D space, the construction process involves the following steps.
1) Division Step: Divide the square into smaller sub-squares. The number of sub-squares depends on the scaling factor \( r \). For example:  If \( r = \frac{1}{2} \), the square is divided into \( 2 \times 2 = 4 \) smaller squares.
2) Random Removal Step: Each sub-square is removed with probability \( p_C \) or kept with probability \( 1 - p_C \).  In this case, \( p_C = 0.8 \), so each sub-square has an 80\% chance of being removed and a 20\% chance of being kept.
3) Iteration: Repeat the process for each remaining sub-square from the previous step. At each iteration, the remaining squares are further divided, and sub-squares are randomly removed or kept,  continuing   until a desired level of detail is reached.   The expected fractal dimension of the random Cantor set in 2-D is:
$D_f = 2+{\log(p_C)}/{\log(2)}$.

Fig. \ref{fig:Cantor} shows  the numerical estimation of the scaling exponents for the 2-D random Cantor set ($p_C = 0.8$ and $r =\frac{1}{2}$,  where $D_f \approx 1.678$) under a wide range of subsampling. 
Panels A–C show the fully sampled Cantor set ($s = 1 $) and two subsampled versions ($ s = 0.5 $ and $ s = 0.05$), respectively. The radial distribution functions $g(r)$ in Panel D and the estimated fractal dimensions in Panel E (for $p_C = 0.8$) exhibit the same robust behavior observed for the Sierpinski gasket: the curves remain nearly indistinguishable across sampling levels, and the estimates based on $g(r)$ are notably more stable than those from box counting.

\begin{figure}[th!]
\centering
\includegraphics[width=1\linewidth]{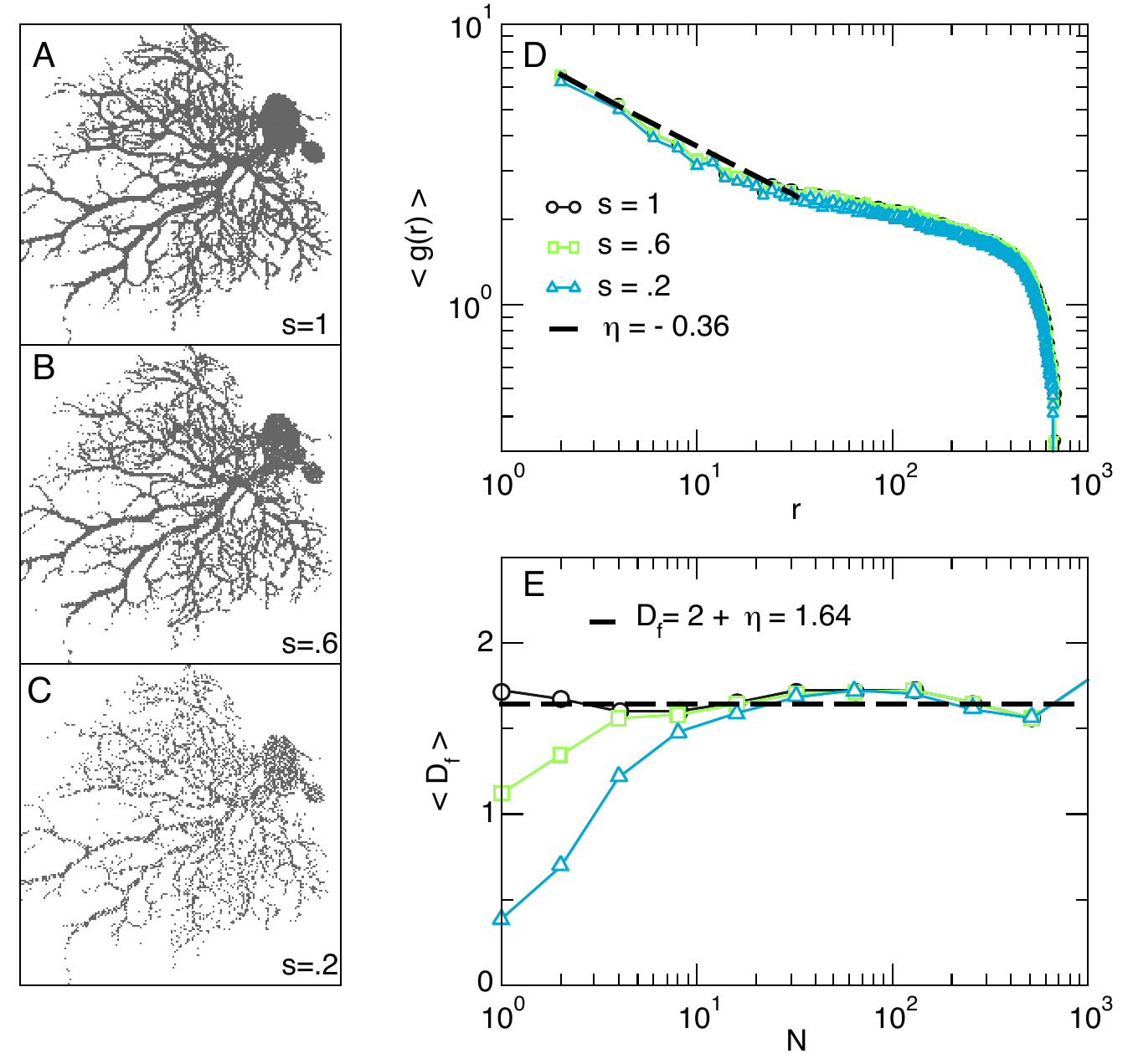}.  
\caption{Maximum intensity projection of a set of optical sections acquired with a multiphoton microscope of a Purkinje neuron from the mouse cerebellar cortex injected with Lucifer Yellow fluorescent dye. Panels on the left shown the fully sampled  binarized image (labeled A), as well as for two different sampling fractions (s=0.6 in panel B and s=0.2 in panel C). Panel D shows the radial distribution function $g(r)$  computed for a range of sampling fractions $s$ and panel E the local fractal dimension for  increasing sampling fractions of the raw binarized image data. Data freely available from \cite{cell}. }
  \label{neuron}
\end{figure}

\textit{1-D time series' correlations behavior under subsampling:}
The correlations scale invariance discussed in the above sections can be defined as well for the case of 1-D time series which is expected to show similar behavior under subsampling. This can be demonstrated by computing the scaling of the fluctuations inside segments of increasing length as implemented by the detrended fluctuation analysis (DFA) \cite{dfa}. 
The DFA method is commonly used to determine the statistical self-affinity of a time series, which may exhibit long range correlations.  The DFA scaling exponent (commonly denoted as $\alpha$) equals the Hurst exponent $H$ \cite{hurst} in the case of stationary processes, but unlike traditional methods, DFA can also assess scaling in non-stationary processes. The relationships among the relevant scaling exponents—the autocorrelation decay exponent $\gamma$, the power spectral exponent $\beta$, $H$, and $\alpha$—can be derived from the Wiener–Khinchin theorem \cite{wiener,WKT,Heneghan}: 
$\gamma=2-2\alpha$;
$\beta=2\alpha - 1$;
$\gamma=1-\beta $;
$\alpha=(\beta+1)/2 $ and
$H = \alpha$ only for fractional Gaussian noise (i.e., for $ -1 \leq \beta \leq 1$).

The relative persistence of the long range correlations under subsampling can be readily demonstrated for 1D time series. This is evident already by simple visual inspection of the time series,  as shown  in panels B and C of Figure \ref{fig:1D} where the overall shape of the signal is preserved, even for sampling rate hundred of times smaller.

\textit{Neuronal structure:}
The correlation behavior under subsampling is now briefly explored  for the scaling of neuronal structures.  The correlation analysis of this type of data can be computationally demanding, especially for high-resolution images which implies the calculation of products of several million pixels. Therefore it is relevant to demonstrate that similar correlation results can be obtained at subsampled images. The results of the analysis is presented in Fig.\ref{neuron}. The fluorescence images were binarized following the methods in  \cite{zamponi2018,zamponi2022} and processed in the same manner as in the synthetic fractal discussed already. It can be seen that $\eta$, the exponent estimated from the initial decay of $g(r)$ is not severely affected by the subsampling  while agreeing with  the expected value for $D_f$. In passing note that the  obtained values are consistent with earlier estimations of $D_f$ for this type of cerebellum neurons reported in Ref.\cite{krauss}.
Box counting here again shows greater sensitivity to subsampling, especially at small box sizes where data sparsity has a stronger effect.
  
 \begin{figure}[th!]
 \centering
 \includegraphics[width=1\linewidth]{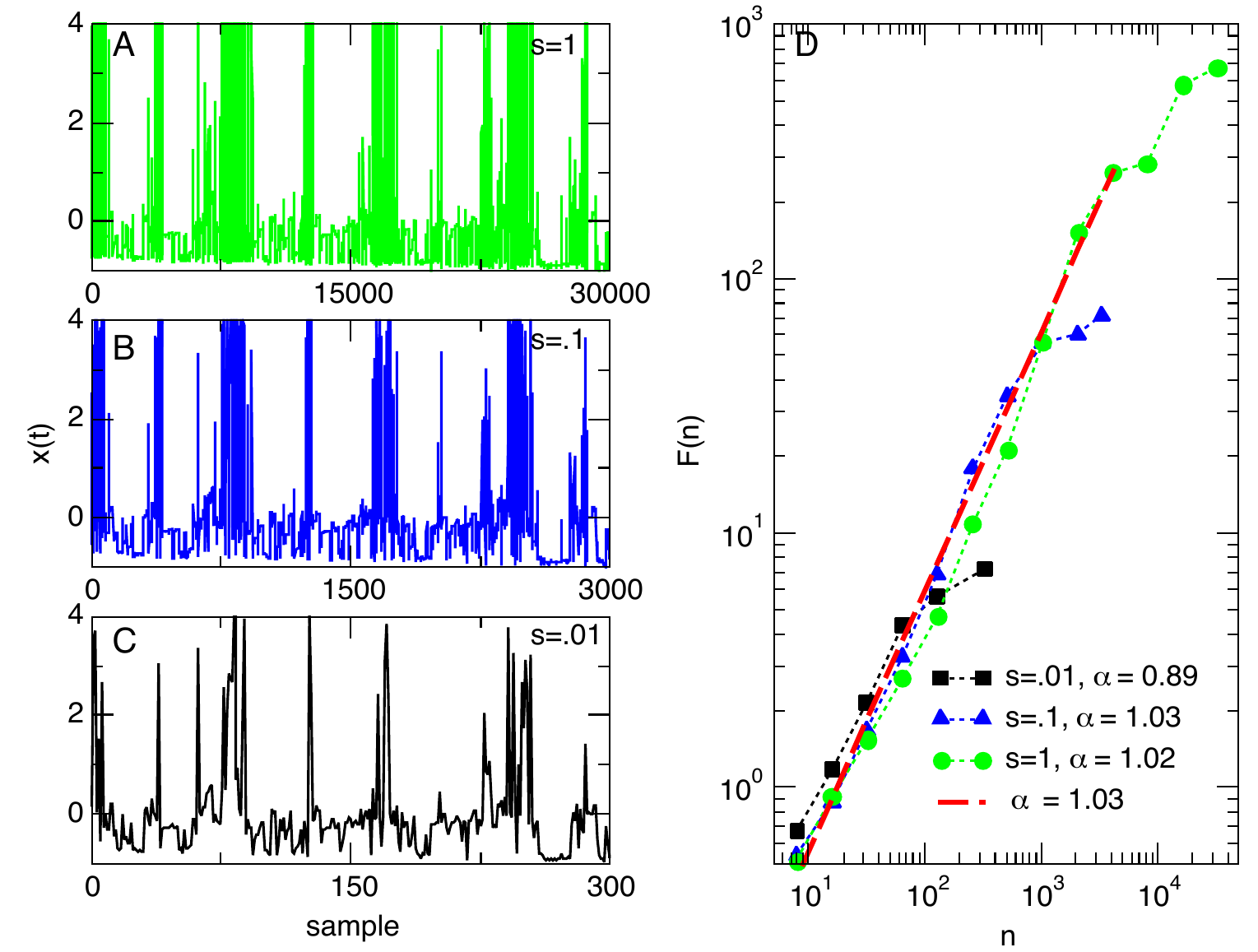}
 \caption{Analysis of longterm records of flies positional tracings. Timeseries shown refers to the position of a single Drosophila melanogaster  inside a thin (3 mm) glass tubes (70 mm length)  registered every 10 seconds for more than 30 days.    Panel  A shows the raw fully sample data. Panels B and C depict the  subsampled versions at  $s=0.1$, and s=$0.01$ respectively and panel D shows the detrended fluctuation analysis of each time series. Data  from Ref.\cite{Geissmann} kindly provided by the authors.}
 \label{fly}
\end{figure}
\textit{Long term correlations in fly's motion behavior:}
Recent studies monitored the movements of fruit fly \textit{Drosophila melanogaster} in an effort to understand how universally conserved phenomenon is sleep among the animal kingdom \cite{Geissmann}. Using machine learning--based video-tracking technology, they conducted a detailed high-throughput analysis of sleep.  To quantify walking dynamics, flies  were introduced on a thin glass tube and loaded into an ethoscope, which is a self-contained machine able to record the activity of flies in real-time using computerised video-tracking with to a resolution of 1,920 x 1,080 pixels, at 30 frames per second (see \cite{ethoscope}  for additional details). 

In Fig.\ref{fly} we analyze one of the data sets  presented in Ref.\cite{Geissmann}.
DFA results show similar behavior to Fig. \ref{fig:1D}: the estimation of characteristic exponent $\alpha$ remains robust even under subsampling by a factor of 100.


\textit{Discussion:} 
Work closely related to the present study merits discussion. Reissa \textit{et al.} \cite{noise_fractals} systematically investigated noise-induced biases in fractal dimension estimation for 2D images, revealing a $\sim 20\%$ inflation of computed values depending on the algorithm and the spatial structure of the fractal. This highlights the sensitivity of local geometric estimators to perturbations, in contrast to correlation-based methods. Another relevant line of work concerns the estimation of critical exponents from avalanche statistics. In systems such as the brain, where only a subset of nodes is typically observable, power-law exponents derived from avalanche size distributions are often biased due to subsampling \cite{ribeiro1,ribeiro2,viola1,viola2,carvalho}. Building on earlier results by Kuntz and Sethna \cite{kuntz}, who showed that the avalanche exponent $\gamma$ equals the spectral exponent $\beta$ in branching processes, Conte and de Candia \cite{Conte} addressed this issue by demonstrating that power spectral and DFA-based exponents remain stable under subsampling. Their findings align with ours, reinforcing the principle that estimators rooted in long-range correlations inherently suppress sampling-related noise.

Summarizing, we examined the effects of uniform stochastic subsampling on the estimation of correlation-based scaling exponents across a range of scale-invariant systems, including 2D spatial patterns, 1D time series (both synthetic and experimental), and biological image data. In all cases, we found that the exponents—whether derived from the initial decay of the radial distribution function $g(r)$ or from DFA—remained remarkably stable under substantial data reduction. This robustness reflects the fact that both subsampling and the estimation procedures suppress short-range fluctuations while preserving the large-scale structure that governs scaling behavior.

Our findings offer practical guidance for empirical studies where full data acquisition is constrained by experimental, computational, or ethical factors. In high-resolution microscopy, large-scale neural recordings, or long-term behavioral monitoring, subsampling is often unavoidable. Yet, our results show that reliable inference of scale-dependent properties remains feasible in such settings—provided the system is statistically homogeneous and suitable estimators are used. This opens the door to principled data reduction without compromising essential structural information.

More fundamentally, such robustness arises from a physical mechanism: both correlation-based estimators and stochastic subsampling act as low-pass filters, preserving long-range structure while suppressing high-frequency fluctuations. This mirrors the aliasing effect in signal processing \cite{oppenheim1999}, where undersampling distorts high-frequency content unless prefiltered. In our case, the smoothing inherent in correlation estimators mitigates such distortions, enabling robust recovery of macroscopic features. This aligns with the universality of scaling laws in critical phenomena, where coarse-graining retains key descriptors despite microscopic variability. A related motivation appears in compressed sensing \cite{donoho2006}, which shows that global structure can be inferred from sparse data when constrained by appropriate priors—though the mechanisms differ.

Notably, a key caveat lies in the requirement of statistical homogeneity: systems exhibiting strong spatial or temporal inhomogeneities may violate the assumptions underpinning our conclusions. While we have focused on scale-invariant correlations—motivated by their widespread relevance across physical and biological systems—similar robustness may extend to short-range correlated systems, provided the correlation length exceeds the subsampling resolution and homogeneity is maintained. More generally, the preservation of macroscopic statistical descriptors under subsampling is expected to depend on the interplay between correlation scale and sampling resolution.

Overall, these insights establish a foundation for structure-preserving subsampling in physical and biological systems. They demonstrate that scaling exponents—far from being fragile quantities—can serve as resilient markers of organization, enabling the analysis of complex systems even under data constraints. As datasets grow in scale and complexity, such principled reduction strategies will be essential for scalable and interpretable scientific insight.

\section{Aknowledgements}
DRC thanks Hong Kong Baptist University for funding his Distinguished Professorship of Science during these studies. 
QYT thanks Natural Science Foundation of China (No. 12305052) and Research Grants Council of Hong Kong (No. 22302723).

 \end{document}